\documentclass[fleqn,10pt,twocolumn]{wlscirep}

\usepackage{graphicx}
\graphicspath{{figures/}}
\usepackage[squaren]{SIunits}
\usepackage{hyphenat}
\usepackage[bookmarks=false] {hyperref}
\usepackage{color}
\usepackage[capitalise]{cleveref}
\crefname{section}{Sec.}{Secs.}
\Crefname{section}{Section}{Sections}
\usepackage{makecell}
\usepackage{physics}

\newcolumntype{L}[1]{>{\raggedright\let\newline\\\arraybackslash\hspace{0pt}}p{#1}}
\newcolumntype{C}[1]{>{\centering\let\newline\\\arraybackslash\hspace{0pt}}p{#1}}
\newcolumntype{R}[1]{>{\raggedleft\let\newline\\\arraybackslash\hspace{0pt}}p{#1}}

\newcommand{\onlinecite}[1]{\hspace{-1 ex} \nocite{#1}\citenum{#1}}

\definecolor{pink}{RGB}{255,0,255}

\definecolor{ss_color}{rgb}{1,0,0}

\definecolor{darkorange}{RGB}{255,120,0}

\definecolor{blue}{RGB}{0,174,179}

\definecolor{darkgreen}{RGB}{20,175,20}

\newcommand{\entmin}[0]{\ensuremath{H_\text{min}}} 
\newcommand{\bitstring}[1]{\ensuremath{\mathbf{#1}}}

\newcommand{\lossPA}[0]{\ensuremath{\text{loss}_\text{PA}}}
\newcommand{\eps}[2]{\ensuremath{\varepsilon^{#1}_\text{#2}}}

\author[1,2,3,4,*]{Shihan~Sajeed}
\author[1,2]{Poompong~Chaiwongkhot}
\author[5,1,3]{Anqi~Huang}
\author[6,1]{Hao~Qin}
\author[7,+]{Vladimir~Egorov}
\author[7]{Anton~Kozubov}
\author[7]{Andrei~Gaidash}
\author[7]{Vladimir~Chistiakov}
\author[7]{Artur~Vasiliev}
\author[7]{Artur~Gleim}
\author[8,2]{Vadim~Makarov}

\affil [1] {Institute for Quantum Computing, University of Waterloo, Waterloo, ON, N2L~3G1 Canada}
\affil [2] {Department of Physics and Astronomy, University of Waterloo, Waterloo, ON, N2L~3G1 Canada}
\affil [3] {Department of Electrical and Computer Engineering, University of Waterloo, Waterloo, ON, N2L~3G1 Canada}
\affil [4] {Department of Electrical and Computer Engineering, University of Toronto, M5S~3G4, Canada}
\affil [5] {Institute for Quantum Information \& State Key Laboratory of High Performance Computing, College of Computer, National University of Defense Technology, Changsha 410073, People's Republic of China}
\affil [6] {CAS Quantum Network Co.,\ Ltd.,\ 99 Xiupu road, Shanghai 201315, People's Republic of China}
\affil [7] {Faculty of Photonics and Optical Information, ITMO University, Kadetskaya line 3/2, 199034 St.~Petersburg, Russia}
\affil [8]{\mbox{Shanghai Branch, National Laboratory for Physical Sciences at Microscale and CAS Center for Excellence in} \mbox{Quantum Information, University of Science and Technology of China, Shanghai 201315, People's Republic of China}}
\affil[*] {shihan.sajeed@gmail.com}
\affil[+]{viegorov@itmo.ru}

\date{\today}
\begin{abstract}
	Although quantum communication systems are being deployed on a global scale, their realistic security certification is not yet available. Here we present a security evaluation and improvement protocol for complete quantum communication systems. The protocol subdivides a system by defining seven system implementation sub-layers based on a hierarchical order of information flow; then it categorises the known system  implementation imperfections by hardness of protection and practical risk. Next, an initial analysis report lists all potential loopholes in its quantum-optical part. It is followed by interactions with the system manufacturer, testing and patching most loopholes, and re-assessing their status. Our protocol has been applied on multiple commercial quantum key distribution systems to improve their security. A detailed description of our methodology is presented with the example of a subcarrier-wave system. Our protocol is a step towards future security evaluation and security certification standards.
\end{abstract}

\title{An approach for security evaluation and certification of a complete quantum communication system}

\begin{document}

	\flushbottom
	\maketitle
	
	\thispagestyle{fancy}
	

\section{Introduction}
\label{sec:introduction}

\noindent
Instead of relying on assumptions of computational hardness like most other classical cryptography protocols, quantum cryptography relies on the laws of physics for providing information-theoretic security. From the first theoretical proposal in 1983~\cite{bennett1984} to the recent key exchange via satellite over $1200~\kilo\meter$ \cite{yin2017}, quantum key distribution (QKD) has come forward a long way. Over the course of time, the journey has been (and is still being) impeded by a number of attacks that exploit the deviations between theory and practice \cite{lydersen2010a,xu2010,sun2011,sajeed2015,sajeed2015a,sajeed2016,makarov2016,huang2016,sajeed2017,pinheiro2018}. Ironically, as a consequence of the attacks, QKD has been equipped with improved protocols and tools like decoy states \cite{hwang2003,lo2005}, measurement device independence \cite{lo2012}, device-independence~\cite{acin2007}, twin-field QKD~\cite{lucamarini2018} and so on. As a result, QKD today is much more secure and efficient in practice than it was 20 years ago.

It is now time for QKD to be expanded and deployed on a larger scale. As the push from the lab to practical deployment is initiated in various parts of the globe, a number of security, compatibility and connectivity issues are needed to be solved. These demand developing universally accepted standards and certification methodologies, and also the formation of a common platform for collaboration and addressing these issues. To fulfil this need, ETSI has had an industry specification group for QKD (ISG-QKD) since 2008 that provides a platform for the creation of universally accepted standards and promotes coordination, cooperation and standardization of research for QKD \cite{langer2009,8-etsi-white-paper,27-etsi-white-paper}. Development of security certification standards is at present being discussed in this group and in other standards organisations such as International Organisation for Standardisation (ISO) \cite{qkdsr-iso} and International Telecommunication Union (ITU) \cite{1-qkdov2019-itu-sg17,2-qkdtn2019-itu-sg17}. At the same time, several recent studies attempt to introduce certification of countermeasures against specific vulnerabilities in a quantum-optical part. For example, Ref.~\onlinecite{lucamarini2015} studies the security of a photon source in a fiber-based QKD system against a general Trojan-horse attack (THA)~\cite{vakhitov2001,gisin2006}. By treating the attack as an information leakage problem, the secure key rate becomes a function of the specifications of the installed optical components. The latter can be characterised when necessary. A similar approach has been suggested for other individual imperfections \cite{xu2015a,pinheiro2018,huang2018,huang2019,wu2020}. A methodology to characterise and secure the source against several imperfections is under development \cite{tomita2019}. Attention to several imperfections and attacks is being paid when designing QKD equipment \cite{dixon2017}.

Although these studies have addressed several individual vulnerabilities, no complete system analysis has yet been reported. This is what we do in this work. We lay out a methodology for security evaluation and certification of a complete quantum communication system against all known implementation imperfections in its quantum optical part. To exemplify how our security evaluation methodology works, we present the results of our initial security evaluation performed at ITMO University and Quantum Communication Ltd.\ (St.~Petersburg, Russia) in 2017. They are therefore the first commercial QKD manufacturer to openly publish the security assessment of their system. We also present the results of follow-up (performed by the manufacturer) to exemplify the follow-up step of our methodology. It consists of theoretical and experimental studies that have allowed the manufacturer to quickly improve implementation security of their product by patching its most prominent loopholes. We hope that our methodology will pave the way for developing security evaluation and certification standards for complete quantum communication systems.

The security evaluation team has performed a very similar initial security evaluation in 2016 on the QKD system Clavis3 from ID~Quantique (Switzerland) and on $40~\mega\hertz$ QKD system from QuantumCTek (China). The follow-up step with the latter is currently in progress. We stress that these two industry projects are highly similar in their methodology, character of results and outcomes to the one reported in this Article. I.e.,\ the methodology we report here is applicable to different commercial systems. However details of vulnerabilities found in the two above-mentioned systems remain confidential at the request of the manufacturers. At the same time, in the case of ITMO's system, the complete security analysis results along with all the vulnerabilities and follow-up tasks have been presented here and no information has been kept hidden.

The Article is organised as follows. Our proposed layered architecture of the complete quantum communication system is presented in \cref{sec:Q} and our severity rating scheme for the implementation imperfections in \cref{sec:C}. We describe the system under test in \cref{sec:system}. Our initial security evaluation results are presented in \cref{sec:vulnerabilities} and the follow-up from the manufacturer is presented in \cref{sec:follow-up}. We conclude in \cref{sec:conclusion}.

\section{Security evaluation and certification methodology}

Our methodology requires an iterative interaction between the manufacturer and the certifiers. The certifying agency needs to have an in-depth knowledge of and physical access to the system in order to perform its security evaluation and certification. Thus, the issue of trust has to be implicit in all the security certification tasks. The first stage in our methodology is the \emph{security evaluation stage} by the testing team; then the \emph{follow-up stage} by the manufacturer; then again the \emph{security evaluation stage,} and so on. Through this iterative process, the system security is gradually expected to reach a level that can be trusted and widely accepted. 

The security evaluation stage consists of: (a)~subdivision of the complete system into seven layers based on the definitions provided in \cref{tab:Q}; (b)~scrutinising the system for implementation vulnerabilities that may make it vulnerable against known attacks, as well as trying to find any new unknown attacks that may apply; and (c)~categorising each discovered vulnerability according to the hardness level defined in \cref{tab:C}. When the evaluation stage ends, the follow-up stage starts. In this stage, the security evaluation results are provided to the manufacturer and the patching commences. We have used this same methodology to evaluate the ITMO's system and the two other systems mentioned above.

\subsection{System implementation layers}
\label{sec:Q}

\begin{table*}
	\caption{{\bf Implementation layers in a quantum communication system.}}
	\label{tab:Q}
	\begin{tabular}[t]{>{\raggedright}p{52mm} >{\raggedright\arraybackslash}p{116mm}}
		\hline\hline
		\bf{Layer} & \bf{Description} \\
		\hline
		\hangindent=4.5ex Q7.~Installation and maintenance & Manual management procedures done by the manufacturer, network operator, and end users.\\
		\hangindent=4.5ex Q6.~Application interface & Handles the communication between the quantum communication protocol and the (classical) application that has asked for the service. For example, for QKD this layer may transfer the generated key to an encryption device or key distribution network. For quantum secure direct communication this layer transfers secret messages from/to an external unit that sends and receives them.\\
		\hangindent=4.5ex Q5.~Post-processing & Handles the post-processing of the raw data. For QKD it involves preparation and storage of raw key data, sifting, error correction, privacy amplification, authentication, and the communication over a classical public channel involved in these steps.\\
		\hangindent=4.5ex Q4.~Operation cycle & State machine that decides when to run subsystems in different regimes, at any given time, alternating between qubit transmission, calibration and other service procedures, and possibly idling.\\
		\hangindent=4.5ex Q3.~Driver and calibration algorithms & Firmware/software routines that control low-level operation of analog electronics and electro-optical devices in different regimes.\\
		\hangindent=4.5ex Q2.~Analog electronics interface & Electronic signal processing and conditioning between firmware/software and electro-optical devices. This includes for example current-to-voltage conversion, signal amplification, mixing, frequency filtering, limiting, sampling, timing-to-digital and analog-to-digital conversions.\\
		\hangindent=4.5ex Q1.~Optics & Generation, modulation, transmission and detection of optical signals, implemented with optical and electro-optical components. This includes both quantum states and service optical signals for synchronization and calibration. For example, in a decoy-state BB84 QKD protocol this layer may include generation of weak coherent pulses with different polarization and intensity, their transmission, polarization splitting and detection, but also optical pointing-and-tracking for telescopes.\\
		\hline\hline
	\end{tabular}
\end{table*}

\noindent
Security analysis of a complete quantum communication system is a complex procedure that requires different areas of expertise. To simplify the job and ensure that people with specific expertise can tackle the right problems, it is necessary to subdivide the implementation complexity into layers. As a first step of our \emph{security evaluation} methodology, we have subdivided the system implementation into seven layers based on a hierarchical order of information flow and control as presented in \cref{tab:Q}. Our layer structure is conceptually similar to the open systems interconnection (OSI) model for telecommunication systems \cite{wikipedia-OSI-model}. Just like OSI layers, a layer in our system serves the layer above it and is served by the layer below; however, unlike OSI, all our layers are inside one system, and most of them are not abstraction layers. When a generic system is installed, it starts with the top layer: Q7 installation and maintenance; then operation and processing is subsequently initiated in each underlying layer until it gets down to handling quantum states in Q1 optics layer. Once the optics layer generates photon detections, they are again processed in each layer above in sequence until the top layers: either Q6 interfacing the output of the quantum protocol with the application that has requested it, or all the way up to Q7. Below we explain the functioning of each layer with examples.
	
The lowest layer Q1 handles the photonic signals that carry the quantum states and service functions. The next layer Q2 interfaces the optical components with digital processing and possibly performs some analog signal processing. It contains analog electronics and digital-analog converters. Q3 comprises digital and software algorithms that immediately control the electronics and optics, including its calibration aspects. It might contain, for example, a set of algorithms to maintain avalanche photodiode (APD) temperature, bias voltage, and gating. The next layer Q4 is software that decides which Q3 layer subroutine to run. For example, it decides when APDs need to be cooled, or when gating control should be initiated. The next layer Q5 processes the raw data generated by the hardware to distill the final data in the protocol, for example generate secret keys in QKD. The layer above Q6 handles the communication between the quantum protocol and the classical application that asks for the service of the protocol. Finally, the topmost layer Q7 handles issues in any underlying layer that require human intervention, even if the human follows a checklist. We have found that the system evaluated in this Article, as well as several other QKD systems, allow a clear division into this layer structure. An example is given in Methods.

We admit that the definition of the layers may not be complete and there could be cases when the functionality of a particular hardware component may span across several layers. In that case, the component accommodates more than one layer. For example, signal processing and algorithms belonging to the layers Q2 through Q5 can be implemented in a single physical field-programmable gate array (FPGA) chip. Also, the ordering of the layers may not be absolute. For example, in some systems, the layer Q5 post-processing may run in parallel with the layers below it while in other systems it may start after the end of Q4 operation cycle. In any case, improvisations have to be made when cutting each system into the layers.

\begin{table*}
	\caption{{\bf Hardness against implementation imperfections.} Here we propose a classification scheme quantifying how robust a given system or countermeasure is against a given imperfection. The hardness level is assigned to each particular imperfection and the same imperfection at different systems may be assigned different levels. For each imperfection the hardness level reflects current knowledge, and may change over time.}
	\label{tab:C}
	\begin{tabular}[t]{>{\raggedright}p{1.5ex} >{\raggedright}p{14.5ex} >{\raggedright}p{55mm} >{\raggedright\arraybackslash}p{78mm}}
		\hline\hline
		\multicolumn{2}{l}{\bf Hardness level} & {\bf Description} & {\bf Examples}\\
		\hline
		C3. & Solution secure & Imperfection is either not applicable or has been addressed with proven security. & The threat of a photon-number-splitting attack on multiphoton pulses is eliminated by the decoy-state protocol \cite{lo2005,hwang2003}; detector imperfections are made irrelevant by measurement-device-independent (MDI) QKD \cite{lo2012}; statistical fluctuations owing to finite sample size are accounted by finite-key post-processing.\\
		C2. & Solution robust & This is the status of many countermeasures after their initial design. With time this state may move up to C3 after a security proof is completed, or down to C1 or C0 after working attacks on it are found. & Phase-remapping in Clavis2 \cite{xu2010} (the imperfection is there, but any known attack attempting to exploit it causes too many errors); long wavelength Trojan-horse attack on Bob in Clavis2 \cite{sajeed2017} (the use of a narrowpass wavelength filter appears to be sufficient given that any known remaining attack causes too many errors).\\
		\hline
		C1. & Solution only partially effective & Countermeasure is successful against certain attack(s), but known to be vulnerable against at least one other attack or a modification of the original attack. & Random-efficiency countermeasure against detector control in Clavis2~\cite{huang2016}; pulse-energy-monitoring system in Alice against Trojan-horse attack~\cite{sajeed2015}; pinhole countermeasure against detector-efficiency-mismatch attacks~\cite{sajeed2015a}.\\
		C0. & Insecure & Security-critical imperfection has been confirmed to exist, but no countermeasure has been implemented. & Laser damage attack on the pulse-energy-monitoring detector in Alice in Clavis2~\cite{makarov2016} and on optical attenuators in several systems~\cite{huang2020}; photon emission caused by detection events in single-photon detectors~\cite{meda2017,pinheiro2018}.\\
		CX. & Not tested & Imperfection is suspected to exist and be security-critical, but has not been tested. & Patch for channel-calibration in Clavis2 \cite{jain2011}; imperfections reported in Ref.~\onlinecite{sajeed2016} against detector-device-independent QKD.\\
		\hline\hline
	\end{tabular}
\end{table*}

If the system contains a separate physical random number generator (RNG), it is considered to be a separate quantum device and therefore not included into our layer classification. Its output would of course interface somewhere with the system, e.g.,\ at layer Q5. Being a separate device it may have an implementation structure of its own, which we do not consider here.

We remark that an initial theoretical proposal of a quantum communication protocol (such as Refs.~\onlinecite{bennett1984,bennett1992b}) covers a part of the single layer Q5, while being mostly ignorant of the other layers except their few selected aspects. However practical security loopholes can be present anywhere in the complete implementation and be in any of its layers. The implementation of each layer has high technical complexity and contain tens of optical components, operator's checklists, thousands of electronic components and lines of software code. The task of security analysis is to find all the loopholes.

\subsection{Quantifying hardness against implementation imperfections}
\label{sec:C}

\noindent
When an implementation imperfection is suspected to be security-critical, it is necessary to evaluate the security risks. The first step is testing. If it is found to be compromising the security then the next step is to design a countermeasure solution, and the last step is checking the robustness of that solution. This procedure is often a loop, because most countermeasures in turn need to be tested. In order to quantify implementation imperfections---existing inside the system---in terms of solutions implemented, we have categorised them as shown in \cref{tab:C}. The lowest state CX indicates that the imperfection is suspected to be a potential security issue, and needs to be further analysed or tested before a conclusion can be made. After an imperfection is found to be security-critical, its state becomes C0, i.e.,\ insecure. Next, a solution needs to be developed that provides security against the original attack model. At this state the solution is expected to be robust and the imperfection is considered to be state C2. After it has been integrated into a security proof, the state can be shifted to C3: solution secure. However, often it may be the case that newer attack models are found that bypass the countermeasure; then the state moves to C1, which means the solution is robust only against a specific attack model but not against others or a combination of the original and some other attacks. 

For example, in ID~Quantique Clavis2 QKD system, the imperfection that the detectors were vulnerable to bright-light detector control attack became C0 upon its discovery in 2009 \cite{lydersen2010a}, was reclassified C2 after being patched in 2015, then downgraded to C1 next year after the patch was demonstrated to be inadequate against a modified attack \cite{huang2016}. A similar development can be traced for another imperfection: variation of detector efficiency with angle of the incoming light \cite{sajeed2015a}. It was suspected to be a security vulnerability (CX) up to 2015, then proven to be so (C0) in 2015 \cite{sajeed2015a,rau2015}, then moved to C2 by the use of a pinhole and later brought down to C1 after the results presented in Refs.~\onlinecite{makarov2016,chaiwongkhot2018}.

We emphasize that the categorisation of a specific vulnerability reflects only the existing knowledge about them which can change with time as seen from the above discussion. Also the categorisation of each existing imperfection depends on the specific system and the specific solution implemented. For example, an imperfection in the single-photon detectors may be classified as insecure (C0) but the same imperfection might be irrelevant (C3) for a system running a measurement-device-independent (MDI) QKD protocol.
 
Eventually, the objective of the security evaluation process should be to upgrade the system such that all imperfections are on the level C3. Level C3 should be considered good for a commercial product, while levels C1, C0 and CX should be deemed inadequate and need to be remedied by a security update or new product development.  Level C2 lies in the gray zone and while it may be considered secure for practical purposes, i.e., adequate for a commercial product, one should remember that it has no theoretical security proof based on quantum mechanics. However, the development of security proofs taking into account imperfections can---in some cases---be a slow process, and we expect many of them to attain C2 earlier than C3.

\subsection{Security evaluation of ITMO's subcarrier-wave quantum key distribution system}
\label{sec:system}

\noindent
In the rest of this Article we demonstrate how our proposed \emph{security evaluation and certification methodology} can be applied to a specific system. As an example, we select the subcarrier-wave quantum key distribution (SCW QKD) system manufactured by ITMO University and its spin-off company Quantum Communications Ltd.\ During the initial security evaluation, the manufacturer has provided us with an overall design specification of the system along with further oral information and written notes on various aspects of design and manufacturing process. We had physical access to the hardware but did not perform any experiments on the setup during that stage. Following the methodology from \cref{sec:Q} and \cref{tab:Q}, we have performed a complete security analysis of the bottom four layers (Q1--Q4) that correspond to optics, analog electronics, driver and calibration algorithms, and operation cycle of the system. For these layers, we have examined all suspected implementation security issues according to the current knowledge. For higher layers Q5 and up (from QKD protocol post-processing and up), we cannot perform a complete security evaluation as they lay outside our expertise area; they should be analysed by a team with expertise in classical information technology security. Nevertheless, we have pointed out a few issues in the layer Q5.

The results of this initial security evaluation have initially been delivered to ITMO in a confidential report in February 2018 (prepared by those authors not affiliated with ITMO). A summary of that report is presented in \cref{sec:vulnerabilities}, after we briefly introduce the system to the reader.

The subcarrier-wave QKD principle was proposed in 1999~\cite{merolla1999-1} and experimentally demonstrated later the same year~\cite{merolla1999}. It was initially conceived as a practical fiber-optic system offering an alternative to then-dominant polarization and time-bin encoding schemes that would require a precise alignment during operation~\cite{merolla1999-1}, as well as to ``plug-and-play'' systems developed a year earlier~\cite{muller1997} that limited QKD source repetition rate due to an intrinsic two-pass architecture. More recently, SCW QKD has been demonstrated as being robust against external conditions affecting the telecom fiber~\cite{gleim2016}, allowing increased spectral density \cite{mora2012,capmany2006}, and being invariant to telescope rotation in open-air links~\cite{kynev2017}. Its viability has been experimentally demonstrated for metropolitan area telecommunication lines~\cite{gleim2017}, multi-user~\cite{chistyakov2014,bannik2017} and software-defined~\cite{chistyakov2017} networks.

\begin{figure}
\centering
	\includegraphics{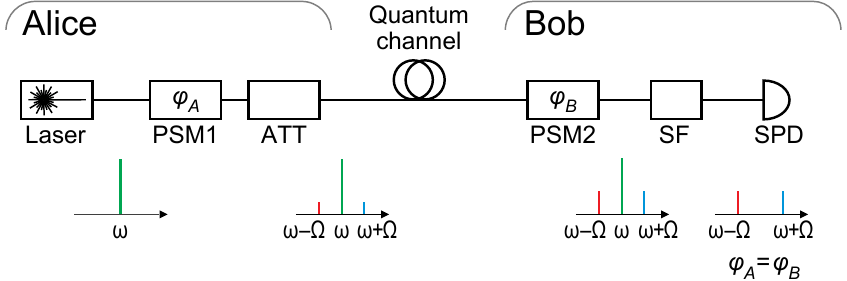}
	\caption{Basic subcarrier-wave QKD scheme. Plots show optical spectra at different points in the setup. ATT,~optical attenuator; PSM,~electro-optical phase shift modulator; SF,~notch spectral filter; SPD,~single-photon detector.}
	\label{fig:SCW-setup}
\end{figure}

A basic design of the SCW QKD system is shown in~\cref{fig:SCW-setup}. In Alice module, a continuous narrow linewidth laser acts as a light source. This radiation with frequency $\omega$ is usually referred as carrier wave, or simply a carrier. It passes through an electro-optical phase modulator, to which an electrical driving signal with frequency $\Omega$ is applied. As a result, two subcarriers (or sidebands) with frequencies $\omega-\Omega$ and $\omega+\Omega$ appear in the optical frequency spectrum, as shown on the plot in~\cref{fig:SCW-setup}. Quantum information is encoded in the phase shift $\varphi_A$ between the carrier and the subcarriers, which is induced by phase modulation of the electrical driving signal~\cite{gleim2016}. Four phase states ($0$, $\pi$/2, $\pi$, 3$\pi$/2) are used in both Alice and Bob modules. After modulation the signal passes to the quantum channel through an attenuator. Carrier power, modulation index and attenuation value are chosen so that the mean photon number $\mu_\text{sb}$ (on two sidebands combined) meets the protocol requirements. On Bob side a similar modulator introduces phase shift $\varphi_B$ resulting in single-photon interference on the sidebands. An optical filter separates the carrier from the sidebands, and the latter are detected on a single-photon detector. The registered optical power depends on the difference $\abs{\varphi_A-\varphi_B}$. If Alice and Bob introduce equal phase shifts, constructive interference is observed, and the optical signal power at the sidebands differs from zero. In the opposite case, when the difference equals $\pi$, destructive interference occurs and the registered counts correspond to dark noise of the detector. Instances when the difference is $\pi/2$ are discarded during sifting. Key bits are obtained from the registered counts using algorithms similar to a phase-encoded BB84 protocol~\cite{miroshnichenko2018,gleim2016}. A full quantum description of the system and the implemented protocols can be found in Refs.~\onlinecite{miroshnichenko2018,kozubov2019}.

\subsection{Potential vulnerabilities }
\label{sec:vulnerabilities}

\begin{table*}
	\caption{{\bf Summary of potential security issues in ITMO's subcarrier-wave QKD system.} $C_\text{init}$, hardness of the initial implementation (analysed in 2017) against this security issue; $C_\text{curr}$, hardness of the current implementation (patched as of early 2020) against this security issue. $C_\text{curr}$ reflects the current knowledge about the security issue, and may change in the future (see \cref{sec:C}). $Q$,~system implementation layers involved (see \cref{sec:Q}).}
	\label{tab:attacks}
\small{
	\begin{tabular}[t]{L{1.8cm}L{0.6cm}L{0.8cm}L{1.5cm}L{1.5cm}L{1.1cm}L{1.4cm}L{0.7cm}L{4.4cm}}
	\hline\hline
	\makecell{Potential\\ security issue} & 
	\makecell{$C_\text{init}$} & 
	\makecell{$Q$} & 
	\makecell{Target\\ component} & 
	\makecell{Brief\\ description} & 
	\makecell{Require \\ lab\\ testing?} & 
	\makecell{Initial\\ risk\\ evaluation} &
	\makecell{$C_\text{curr}$} & 
	\makecell{Current status} \\
	\hline
	
	Controllable detectors & CX & Q1--5,7 & SPDs & See Ref.~\onlinecite{lydersen2010b}. & Yes & High & C2 & Loophole has been experimentally confirmed and the suggested countermeasures~\cite{chistiakov2019} have been implemented in the current version. \\
	
	Laser damage & CX	 & Q1,3 & Alice's \& Bob's optics & See Ref.~\onlinecite{makarov2016}. & Yes & High & C2 & Loophole has been experimentally confirmed in Alice and the suggested countermeasures~\cite{huang2020} have been implemented in the current version. \\
	
	Trojan horse & C2, \newline C0 & Q1 & Alice's \& Bob's optics & See Ref.~\onlinecite{lucamarini2015}. & Yes & Low (Alice), High (Bob) & C2, \newline C2 & Manufacturer has developed countermeasures (patent pending) to be implemented in the next system modification and then analysed again by the testing team. \\
	
	Lack of general security proof & C0 & Q1,5 & QKD protocol & \cref{sec:reference-monitoring}. & No & High & C3 & Was a known issue. Has been covered by the manufacturer after receiving the report, see Ref.~\onlinecite{kozubov2019}. The privacy amplification procedure has been updated in the software. The two groups continue to jointly verify the security proof. \\
	
	Manipulation of reference pulse & CX & Q1,5 & QKD protocol & \cref{sec:ref-pulse}. & No & High & C3 & Was a known issue. Has been covered by the manufacturer after receiving the report, see Ref.~\onlinecite{kozubov2019}. Reference monitoring has been implemented in the system. \\
	
	Time-shift attack & CX & Q1--3,5 & PSMs & \cref{sec:time-shift}. & Yes & Medium  & CX  & Lower priority issue that is a subject for future work. \\
	
	Privacy amplification & C0 & Q5 & Post-processing & \cref{sec:privacy-amplification}. & No & High & C3 & Was a known issue. Has been covered by the manufacturer after receiving the report, see Ref.~\onlinecite{kozubov2019}. The privacy amplification procedure has been updated in the software. \\
	
	Finite key size effects & C0 & Q5 & QKD protocol & See Ref.~\onlinecite{tomamichel2012}. & No & Low & C3 & Was a known issue. Has been covered by the manufacturer after receiving the report, see Ref.~\onlinecite{kozubov2019}. The system software has been updated taking the finite-sized effects into account. \\
	
	Non-quantum RNG & C0 & Q5 & RNG & \cref{sec:rng}. & No & Low & C3 & Was a known issue. The manufacturer has put effort into quantum RNG research~\cite{ivanova2017fiber} and has selected a physical RNG for the next version of the system. \\
	
	Intersymbol interference & CX & Q1--3 & PSM's drivers & \cref{sec:intersymbol-interference}. & Yes & Low & CX & Lower priority issue that is a subject for future work. \\

	\hline\hline
	\end{tabular}
	}
\end{table*}

\noindent
Based on the received information about the system, we have identified a number of potential security issues that might be exploitable by an adversary Eve. A summary of these results is given in \cref{tab:attacks}. For each imperfection, we specify the corresponding $Q$-layers (see \cref{sec:Q}), hardness level $C_\text{init}$ (see \cref{sec:C}) and an estimate of the risk. Almost all the identified issues require further detailed analysis, and in many cases, in-depth experimental testing in a laboratory. For many issues, the hardness  level is CX, meaning the issue's applicability to the system implementation needs to be studied and tested. We specify in which system implementation $Q$-layers each issue is located, according to the classification introduced in \cref{sec:Q}.

The risk evaluation listed in \cref{tab:attacks} is based on a guessed likelihood of the vulnerability, expected fraction of the secret key leakage, and estimated feasibility of exploit technology. It is essential for manufacturers with limited resources to prioritize the problems. Vulnerabilities that can be exploited using today's technology and compromise full secret key are a more immediate threat. They should be addressed before those that require future technology or provide only partial key information (thus requiring of Eve an additional classical cryptanalytic task). We have followed this strategy and tested the two highest risk issues during the follow-up stage (see \cref{sec:follow-up}). The security proof and implementation of post-processing have also been completed after the report.

We remark that more security issues may be discovered in the future once the system design and operation are examined in greater detail. We now explain the identified issues.

\subsubsection{Controllable detectors}
\label{sec:detector-control}

\noindent
Two types of detectors are used in the present implementation: ID~Quantique (IDQ) ID210 gated APD and Scontel~TCORPS-CCR-001 superconducting nanowire single-photon detector (SNSPD). Among them, Scontel SNSPD is at least partially controllable by bright light~\cite{lydersen2011c,tanner2014,elezov2019}. Whether the same was true for ID210, required experimental testing. From our previous measurements on ID~Quantique Clavis2 QKD system, we know that it is possible to blind its detectors by sending a continuous-wave (c.w.)\ light of power $P_\text{blind} = 0.3~\milli\watt$~\cite{huang2016}. Then by choosing a trigger pulse power $P_\text{tr}$ greater than the threshold power $P_\text{th} = 0.15~\milli\watt$, it is possible to force a click when Bob-Eve phases match. If we assume ID210 behaves similarly to the detectors in Clavis2 system, then Eve could send c.w.\ power to blind it and perform the faked-state attack~\cite{lydersen2010a} detailed in Methods.

However, sending a trigger power $P_\text{tr}$ at the subcarrier frequency will not work as the photons will be shifted to another frequency due to Bob's modulation. Instead, Eve needs to inject extra photons in the reference signal frequency so that they are shifted to the subcarrier after the modulation and trigger a click in the blinded detector. Due to the small $m$ in the present system, the reference power required by Eve is $P_\text{ref} \approx P_\text{tr} / m$. For example, for $m = 0.05$, a $1~\nano\second$ trigger pulse at the subcarriers with peak power $P_\text{tr} > 0.15~\milli\watt$~\cite{huang2016} just before the detectors would require a $1~\nano\second$ wide reference pulse with peak power of $P_\text{ref} > 3~\milli\watt$ at Bob's input. This is an easily generated and transmitted optical power.

\subsubsection{Laser damage}
\label{sec:lda}

\begin{figure}
	\centering
	\includegraphics{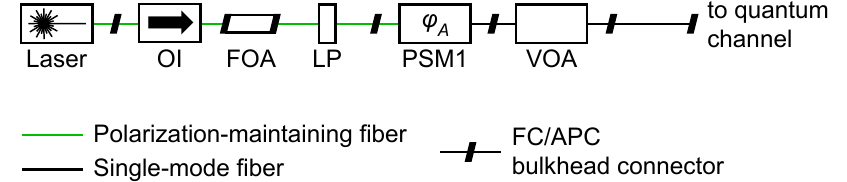}
	\caption{Alice's optical scheme in detail. Component pigtails are connected using angled ferrule connectors (FC/APC). OI,~optical isolator; FOA,~fixed optical attenuator (plug-in style); LP,~linear polarizer; VOA,~variable optical attenuator.}
	\label{fig:SCW-alice}
\end{figure}

\noindent
Whether the current system is vulnerable to laser damage attack (LDA)~\cite{bugge2014,makarov2016}, can be ascertained only after experimental testing. Since one of attenuating components, a variable optical attenuator (VOA; FOD 5418) in Alice is the closest to the channel (see \cref{fig:SCW-alice}), it will be the first target for Eve's LDA. Eve can send high power laser to damage the optical attenuator to reduce its attenuation. If successful, lights coming out of Alice will have higher mean photon numbers than permitted by the security proofs, thus compromising the security.

It will also be interesting to experimentally check the effect of laser damage on the optical PSMs to see whether LDA can affect $m$. If it can, then further studies need to be conducted to check whether it leads to a denial of service or a security compromise. Finally, if LDA can reduce the insertion loss of either the PSM1, linear polarizer (LP) or fixed optical attenuator (FOA) in Alice, it may facilitate other attacks, e.g.,\ Trojan-horse attack. Hence, these components must be characterized meticulously against LDA.

\subsubsection{Trojan-horse attack}
\label{sec:Trojan-horse}

\noindent
In SCW QKD protocol, after sifting, Alice and Bob keep only the outcomes for which they both used the same phase, i.e., $\varphi_A = \varphi_B$. Thus if Eve can extract information on either $\varphi_A$ or $\varphi_B$ by performing a Trojan-horse attack (THA)~\cite{vakhitov2001,gisin2006,jain2014}, the security will be compromised. With current technology, Eve needs a mean photon number $\mu_{B\rightarrow E} \sim 4$ to perform homodyne detection~\cite{jain2014}.

The secure key rate in the presence of THA---under reasonable assumptions---is available for both single-photon and decoy-state Bennett-Brassard 1984 (BB84) protocol \cite{lucamarini2015}. It is based on Alice's ability to upper-bound the outgoing mean photon number $\mu_\text{out}$. A similar theoretical analysis under assumptions appropriate for the present scheme is not available, and needs to be performed. Moreover, wavelength can also be an attack variable \cite{jain2015,sajeed2017}. It is thus important to measure experimentally the actual values of the insertion loss and reflection coefficients of several components such as LP, FOA, OI, connectors, etc.\ in a large range of wavelengths that can propagate through the optical fiber (from $< 400$ to $> 2500~\nano\meter$). Since a laboratory with wideband characterisation equipment is not readily available to us, we have limited our analysis to Eve using a single $1550~\nano\meter$ wavelength. With these two shortcomings, our security evaluation of the system against the THA is detailed below.

\textbf{Alice:}
In the present scheme (\cref{fig:SCW-alice}), possible sources of reflection are the LP (Thorlabs ILP1550PM-APC), FOA (Fibertool FC-FC 15~dB), optical isolator (OI; AC Photonics PMIU15P22B11), all the standard optical connectors placed after PSM1 (i.e.,\ at its side facing away from the quantum channel), and that facet of PSM1. We identify that one of the strongest sources of reflection is the LP with $45~\deci\bel$ return loss (according to its data sheet). Assuming the VOA is set to $70~\deci \bel$ (which is a typical attenuation value required by the SCW QKD protocol), the insertion loss of the PSM1 is $3~\deci\bel$ and that of each connector is $0.3~\deci\bel$, the total round-trip attenuation experienced by a Trojan photon is $193.4~\deci\bel$. For the other protocols, an appreciable decline of performance begins at $\mu_\text{out} \sim 10^{-6}$ \cite{lucamarini2015}. For that, an eavesdropper would need to send $2.2 \times 10^{13}$ photons per pulse into the system, which---considering a phase change frequency of $f = 100~\mega\hertz$---corresponds to injecting c.w.\ power of $280~\watt$. This is somewhat above present-day technology capability, may be around the physical limit of how much power the standard fibers can carry, and will certainly trigger laser damage of Alice's components. Most fiber-optic components get damaged at less than $10~\watt$ \cite{makarov2016,huang2020,ponosova2020,ruzhitskaya2020}. While this suggests the risk of THA at Alice's side is relatively low, it is important to check the reflection from the OI and FOA, which requires experimental testing. Finally, this analysis should be repeated for lower attenuation settings of the VOA that may be used by the system and the risk should be evaluated accordingly.

\begin{figure}
	\centering
	\includegraphics{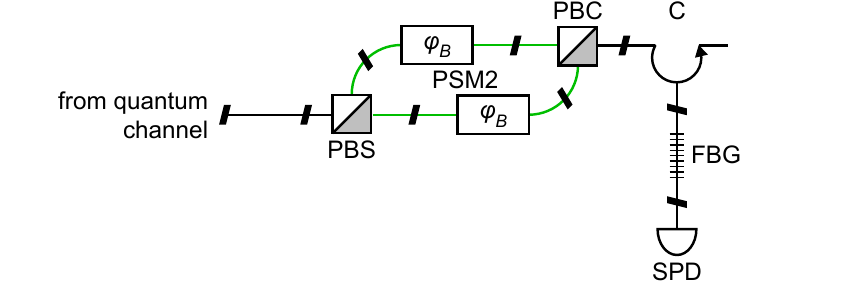}
	\caption{Bob's optical scheme in detail. Bob's phase shift modulator PSM2 is polarization-insensitive and is implemented as two identical modulators acting on orthogonal components of input polarization. PBS,~fiber-optic polarization beam splitter; PBC,~fiber-optic polarization beam combiner; C,~circulator; FBG,~fiber Bragg grating.}
	\label{fig:SCW-bob}
\end{figure}

\textbf{Bob:}
The risk of THA on Bob seems to be comparatively higher than that at Alice since there is no attenuator or isolator in Bob's module (\cref{fig:SCW-bob}). The reflection coefficient of the polarization beam combiner (PBC; AC~Photonics PBS15P12S11-2m) just after PSM2 is $50~\deci\bel$ (according to its data sheet) while the insertion loss of the polarization beam splitter (PBS; same as PBC), PSM2, and each of the four connectors is $0.48$, $1.7$, and $0.3~\deci\bel$, respectively. Assuming the point of reflection is the PBC just after the phase modulator, the total loss experienced by a Trojan photon will be $l =56.8~\deci\bel$. This means that in order to get a single photon out, Eve needs to inject a c.w.\ power of only $6~\micro\watt$, which is easy.

Note that ID210 runs in gated mode with afterpulsing. So, Eve can send the Trojan photons just after the gate but still inside the phase modulation window. However, this may cause a high level of afterpulsing in Bob's single-photon detectors~\cite{jain2014}. Scontel TCORPS-CCR-001 has no afterpulsing but it runs in continuous mode, thus making it difficult for Eve to send Trojan photons. Eve can resort to a longer wavelength (such as $1924~\nano\meter$) to reduce both the afterpulsing side-effect~\cite{sajeed2017} and also the probability of the Trojan photons to be detected. As a result, wavelength filters are necessary in Bob. Nevertheless, afterpulsing characterization of detectors along with characterization of the wavelength filter at longer wavelengths are necessary in order to prevent the THA.

\subsubsection{Lack of general security proof}
\label{sec:reference-monitoring}

\noindent
An apparent requirement of the SCW QKD protocol (to prevent photon number splitting (PNS) ~\cite{bennett1992b,lutkenhaus2002} and unambiguous state discrimination (USD) attack~\cite{dusek2000,sajeed2015}) is to monitor the carrier signal as highlighted in \cite{bennett1992c,merolla1999}. However, based on our discussions with ITMO's engineers, we learned that the monitoring of the reference signal might not be implemented partly due to implementation complexity and partly because they do not deem it necessary for security, because Ref.~\onlinecite{miroshnichenko2018} shows that the system is secure against a collective beam splitting (CBS) attack over a large distance. Here, we emphasize that being secure against the CBS attack mentioned in Ref.~\onlinecite{miroshnichenko2018} does not guarantee security against more general attacks. As an example, we outline a more powerful attack in Methods. 

\subsubsection{Manipulation of reference pulse}
\label{sec:ref-pulse}

\noindent
Here we assume that the reference pulse monitoring is implemented in the system and analyse the consequences. If care is not taken during the implementation, there might still be ways for Eve to perform the USD attack as the following. 

First, Eve intercepts Alice's signal just outside Alice's lab and performs a USD measurement~\cite{huttner1996}. For any conclusive measurement, she prepares the same state with a higher mean photon number and sends it to Bob via a lossless channel, in order to maximize his detection probability. For any inconclusive measurement, she still needs to send the reference signal to Bob and wants it to be detected. However, sending only the reference signal while suppressing the sidebands does not work as it will introduce errors. Instead, Eve wishes the subcarrier signal detection probability to be as low as possible while still keeping the reference signal detection probability as high as possible.  The number of photons in the subcarrier and reference signal---after Bob's modulation---is given in Ref.~\onlinecite{miroshnichenko2018} as
\begin{equation}
	\begin{aligned}
		n_\text{ph}^\text{sb} &=\alpha \mu_0 \eta(L) \eta_B \left[ 1 - |d_{00}^s (\beta')|^2 \right],\\
		n_\text{ph}^\text{ref} &=\alpha \mu_0 \eta(L) \eta_B |d_{00}^s (\beta')|^2.
	\end{aligned}
\end{equation}
Here, $\mu_0$ is the mean photon number of the reference pulse, $\eta(L)$ is channel transmission, $\eta_B$ is transmission in Bob module, and $\alpha$ is additional loss induced by Eve. $ |d_{00}^s (\beta')|$ is the Wigner $d$-function that decides the number of photons to be shifted from reference to side-bands based on its argument $\beta'$, which itself is a function of the modulation index and the phase difference between Alice and Bob. 

We assume APDs are used for the detection of both the reference and subcarrier signals. Then the detection probability in mode $i \in \{ \text{ref, sb}\}$ is $P_\text{det}^i = 1 - e^{-n_\text{ph}^i}$ (for simplicity, we consider unity detection efficiency). For normal operation, $n_\text{ph}^\text{sb} \ll n_\text{ph}^\text{ref}$, which leads to $P_\text{det}^\text{sb} \ll  P_\text{det}^\text{ref}$. Depending on the chosen value of $m$ and $ \mu_0$, $P_\text{det}^\text{sb}$ can be significantly more sensitive to $\alpha$ compared to $P_\text{det}^\text{ref}$. In that case, increasing $\alpha$ would reduce $P_\text{det}^\text{sb}$ much faster than $P_\text{det}^\text{ref}$. As a result, it might be possible for Eve to reduce subcarrier signal detection rate without affecting the reference detection rate considerably. The small reduction in $P_\text{det}^\text{ref}$ can be compensated by adjusting the power of the pulses sent during the conclusive measurement cases. The only limitation on $\alpha$ is that $P_\text{det}^\text{ref}$ should not be lowered significantly for Alice and Bob to notice. A countermeasure to this attack can be to monitor the reference and subcarrier detection rates. However, a further study is required to find the optimal strategy to monitor the reference and subcarriers and also to design the monitoring detector, determine $\mu_0$, monitoring threshold, and $m$. 

\subsubsection{Time-shift attack}
\label{sec:time-shift}

\noindent
In order to achieve time synchronization, Alice sends to Bob a continuous 10~MHz sinusoidal optical signal, which is further modulated by a signal of a special shape with $60~\milli\second$ period. The position of bit slots of $10~\nano\second$ period \cite{gleim2016} and other time intervals are defined with respect to this signal. We suspect that it might be possible for Eve to control the time delay of the reference and side-band signals relative to this synchronization signal to shift their arrival times into a specific moment inside or outside the phase modulation window. This might make the system vulnerable against time-shift attacks (TSA)~\cite{zhao2008}. A time-shift attack can be performed on the SCW QKD system as follows. For ease of understanding, let us first assume that there is a time gap between successive phase modulation windows (i.e.,\ they are narrower than the bit slot), and in between the modulation windows the phase is $0$. We assume a faked-state attack in which Eve stays outside of Alice's module and performs USD of Alice's states. Whenever she obtains a conclusive outcome, she sends the same state $\varphi_E$ to Bob in the correct time window (i.e.,\ she does not alter the arrival time). When Bob measures in the same basis, and $\varphi_E = \varphi_B$ ($\varphi_E \ne \varphi_B$), he gets a click (no click). However, when Eve obtains an inconclusive outcome, she generates a $\varphi_E = \pi$ state and sends it in-between the phase modulation windows. Since in between the modulation window the phase applied is $0$, this ensures no detection by Bob's detector.
	
In our discussion with the developers, we learned that in the current SCW QKD implementation, there is no gap between successive phase modulation windows. However, at the transition region from one window to the next, there is a fast fluctuation. Thus, it will be interesting to know what effective phase shift is experienced by a pulse if it is sent at the time interval corresponding to the fluctuations. For example, if the effective phase shift is $\varphi_0$, then it might still be possible for Eve to remain inconspicuous during the inconclusive measurement slots by sending a state $\varphi_E = \pi + \varphi_0$. However, the feasibility of this attack can only be ascertained by experimental testing. For that, one needs to characterize Bob's phase modulation windows---including the transition regions---in the time domain for all phase values. Click processing by Bob will also need to be checked for detection times in the transition regions.

\subsubsection{Privacy amplification method}
\label{sec:privacy-amplification}
	
\noindent
In the composability framework of QKD \cite{renner2005b}, to achieve $\epsilon$-security, it is required that Alice and Bob estimate the upper bound of Eve's information on their key up to the end of error correction step, and apply a proper universal-2 hash function. This is done to generate a shorter secret key such that the probability that the key is not perfect and the protocol did not abort is bounded by $\epsilon$. However, the present system does privacy amplification by first calculating secret key size and then randomly discarding bits in the error-corrected key to match that calculated secret key size. The disadvantage of this random key removal procedure compared to hashing is that Eve can listen to the classical communication between Alice and Bob and follow the exact procedure to discard bits from her own set. At the end, $\epsilon$-security cannot be guaranteed. To make the secret key $\epsilon$-secure according to the composability framework, the proper implementation of privacy amplification using the hash function is advised.

\subsubsection{Finite-key-size analysis}
\label{sec:fk}

\noindent
In the present system, the size of the raw key is limited by the size of Alice's memory (1~Mbit). According to the developers, this leads to a sifted key size of $\approx 20~\kilo$bit for a distance of 12~km. For a larger distance of $200~\kilo\meter$, the size becomes as low as $\approx 10~\kilo$bit~\cite{gleim2016}. $10\%$ of this sifted key is used for parameter estimation. This small sample size has a high probability to lead to discrepancies between the estimated and actual parameter values due to finite-size-effects~\cite{ben-or2005}. Since the present security proof used by the developers does not consider the finite-key-size effects, the system might be vulnerable to them.

Based on our previous analysis on a different system~\cite{chaiwongkhot2017}, we know that the finite-size effects become significant when the sifted key size is lower than $200~\kilo$bit. At that size of the sifted key, the system---without finite-size-analysis---generated a larger secret key than the upper-bound set by the finite-key-size analysis. Thus, security of the generated key was not guaranteed.  Since the sifted-key size of $20~\kilo$bit in the present system is much lower than $200~\kilo$bit, we strongly suspect that finite-size effects are significant. Thus, we advise to develop a thorough finite-key analysis. To do this, any deviation of parameters due to finite-size-effect needs to be analysed. An example of this effect is the collision probability, i.e.,\ the probability of a hash function mapping two different input keys to the same output key. Other examples could be found in Refs.~\onlinecite{cai2009,renner2005a,renner2005b,scarani2008a,tomamichel2012}.

\subsubsection{Non-quantum random number generator}
\label{sec:rng}

\noindent
In the present system, three types of RNGs can be used in an interchangeable manner. One is a pseudorandom number generation software $drand48\_r$ from Linux operating system. The second is a commercial product manufactured by the developers of this QKD system. The third one is the internal RNG of Altera Cyclone~IV FPGA chip. Using a pseudorandom generator (or randomness expansion) does not satisfy the randomness assumption of the security proof. For the other two generators, care should be taken to verify the quantum origin of the random numbers and the quality of implementation.

\subsubsection{Intersymbol interference}
\label{sec:intersymbol-interference}

\noindent
Owing to the limited bandwidth of the driving electronics, high speed systems might exhibit intensity correlation among the neighboring pulses---an effect known as the intersymbol interference or the pattern effect \cite{yoshino2018,roberts2018}. The electronic signal applied to the modulator might be dependent on the preceding pulse, which violates the assumption of security proof. This may lead to vulnerability. Testing should be done in order to assess the risk of the intersymbol interference in the present system.

\subsection{Follow-up stage}
\label{sec:follow-up}

\noindent
After the initial security evaluation report had been delivered in 2017, the follow-up process ensued. Till now, laboratory testing of the two issues \emph{controllable detectors} and \emph{laser damage} has been carried out. In both cases, the testing has confirmed the vulnerability's presence and the manufacturer has designed countermeasures and implemented them in the current version of the SCW QKD system. Most other issues \emph{(Trojan-horse attack, lack of general security proof, manipulation of reference pulse, privacy amplification, finite key size effects, non-quantum RNG)} have also been addressed as outlined below. Two lower-risk issues, \emph{time-shift attack} and \emph{intersymbol interference,} remain to be studied in the future.

\textbf{Controllable detectors:}
Both detector units mentioned in \cref{sec:detector-control} have been tested. It has been found that ID210 is fully controllable by bright light \cite{chistiakov2019}, while Scontel SNSPD with a built-in electronic countermeasure (recently developed by Scontel) is partially controllable and the countermeasure in it needs to be improved \cite{elezov2019}. The optical power required to control ID210 can easily be generated and transmitted through Bob's optical scheme \cite{chistiakov2019}, confirming our original risk assessment. Technical countermeasures against this attack are currently under consideration. We remark that this vulnerability remains unsolved in most existing QKD systems \cite{fedorov2019}.

\textbf{Laser damage attack:}
as suggested in \cref{sec:lda}, we have performed laboratory testing of the VOA unit (FOD 5418). We have found it to be severely vulnerable to the LDA \cite{huang2020}. A brief application of $\sim 2.8~\watt$ c.w.\ laser power damages a metal film layer inside this component and reliably reduces its attenuation by $\sim 10~\deci\bel$, which renders the key insecure. A countermeasure currently under consideration is to insert another component between the line and the VOA, in order to prevent the latter from being exposed to high power. Candidates for this other component are being tested \cite{ponosova2020,ruzhitskaya2020}.

\textbf{Protocol-related issues:}
A proof of security for a general attack---the lack of which has been highlighted in \cref{sec:reference-monitoring}---has been developed in Ref.~\onlinecite{kozubov2019}. It is summarised in Methods. The issues discussed in \cref{sec:ref-pulse,sec:reference-monitoring} have been closed by an analysis of advanced attack and appropriate countermeasures \cite{gaidash2019methods,gaidash2018overcoming}. We recap these results in Methods. Finally, a correct privacy amplification method (\cref{sec:privacy-amplification}) and finite-key analysis (\cref{sec:fk}) have been included in Ref.~\onlinecite{kozubov2019}. The finite-key analysis is recapped in Methods. Since all these issues appear to have been addressed by this recently published theoretical work, we have updated their current hardness level in \cref{tab:attacks} to C3.

Two more issues have also been analysed and patched by the manufacturer. For the Trojan-horse attack (\cref{sec:Trojan-horse}), additional components have been added to the optical scheme in order to detect the attack (patent pending). Also, possible Eve's information acquired by Trojan-horse attack has been quantified and considered in the security model. The non-quantum RNG (\cref{sec:rng}) will be replaced in the next version of the system by a quantum one developed by the ITMO team.

Overall, our joint work has allowed ITMO University and Quantum Communications Ltd.\ to quickly patch most of the loopholes by introducing countermeasures. The implementation hardness levels have been raised from $C_\text{init}$ of CX and C0 at the time of the initial report to the current state $C_\text{curr}$ of mostly C2 or even C3. Countermeasures marked C2 may eventually become C3, after additional experimental testing and improvement. The two groups also continue to jointly verify the protocol security proof.

\section{Conclusion}
\label{sec:conclusion}

\noindent
The lack of security certification methodology for quantum cryptography is ironic, since security is the main concern behind the shift from classical to quantum cryptography. In this work we have presented a methodology for security evaluation of a complete quantum communication system. Our methodology works in an iterative interaction between the certifiers' \emph{evaluation} stage and the manufacturer's \emph{follow-up} stage. At the evaluation stage, the complete system implementation is subdivided into seven layers, a set of layers (in our case the bottom four) are exhaustively searched for vulnerabilities, and finally each discovered imperfection is categorised based on the hardness of the realised solution and practical risk. At the follow-up stage, work is performed to eliminate these vulnerabilities.

We have applied this methodology to three different QKD systems and presented here the results for the SCW QKD system from ITMO University and Quantum Communications Ltd. In this system, we have found a number of potential security issues---which we expose here without omissions---that need a careful investigation by the manufacturer. Experimental tests, countermeasure and theory development have followed. As the result, most of the issues have been addressed, increasing the hardness rating of this implementation. Projects of a very similar character are going on with the two other systems (by ID~Quantique and QuantumCTek) that we earlier analysed. I.e.,\ our protocol is applicable beyond the system detailed in this Article. We hope it will pave the way towards development of a security certification methodology for existing and future quantum communication systems.

Our security certification methodology is developed with only point-to-point QKD protocols in mind and we are not sure how applicable it will be for a network scenario. We hope that making the point-to-point systems secure would eventually make the resultant network secure.

One important but sometimes overlooked aspect should be emphasised. When someone is engaged in designing a system, his mindset tends to become biased, and he may not be able to think from a different point of view and see security problems with his own design. This is the very reason the task of security certification should be done in collaboration with third-party experts whose main goal is to find problems. This helps a responsible QKD manufacturer to quickly assess and resolve the security issues, as has clearly happened in the case of ITMO. Furthermore, the third-party analysis should ideally begin during initial design considerations, rather than after the commercial implementation has been completed (as has been the case here).

\section*{Methods}

\section*{Another example of layer subdivision}
\label{sec:layer-subdivision-example}

\noindent
To give another example, let's consider commercial QKD system Clavis3 \cite{idqclavis3specs}. Its operation can be divided into our proposed layer structure as follows. When a customer receives the system, the first steps involve a manual installation procedure that is done according to the instruction from the manufacturer. For example, the user needs to connect Alice and Bob QKD stations with a fiber, setup two control PCs (running Linux OS) to install the `Clavis3 Cockpit' software, configure an Ethernet network with specific IP addresses to establish communication between control PC and Alice-Bob QKD stations, and connect fibers in Bob QKD station in a specific way depending on whether internal or external single-photon detectors are used. During the course of operation, manual interventions may be needed from time to time for maintenance: for instance, if the control software hangs, a manual restart is required. All these fall under layer Q7. Next, the system should interact with some external key management system or encryption engine. These tasks are handled in layer Q6. Next, layer Q5 specifies the post-processing rules: for example, coherent-one-way (COW) QKD protocol with LDPC error correction (with a code rate $2/3$) and security parameter of $\epsilon = 4 \times 10^{-9}$. Next layer Q4 decides which subroutine to initiate: for instance, whether to adjust synchronization between the Alice and Bob QKD stations, optimise modulator voltages in order to maximize the interference visibility, or send qubits from Alice to Bob. The control is then transferred to layer Q3, which executes the chosen subroutines with help from Q2 and Q1. For example, when Q3 initiates the raw key exchange subroutine, the field-programmable gate array (FPGA) chip in Alice---at layer Q2---outputs a stream of 1.25~Gbps digital pulses with adjustable amplitude and width to drive an intensity modulator that prepares the quantum signals. The latter are then sent over the fiber to Bob. Another FPGA at Bob---a layer Q2 device---outputs another stream of 1.25~Gbps pulses to provide the gating signals to the single-photon detectors and receives detection signals from these detectors. Here, the intensity modulators, fiber, and detectors all belong to layer Q1 that---together with components from layer Q2---executes a subroutine initiated by layer Q3.

\section*{Faked-state attack strategy}
\label{sec:faked-state-attack}

\noindent
Let's assume first that there is no reference monitoring implemented in the system. Let's assume Alice encodes phase $\varphi_A$. We further assume that Eve---sitting outside Alice's module---measures the signal (using similar measurement setup as Bob) by randomly applying $\varphi_E \in \{0, \pi/2, \pi, 3\pi/2 \}$. Another part of her---sitting near Bob---sends bright c.w.\ light of power $P_\text{blind}$ to blind Bob's side-band detector. When $\varphi_E = \varphi_A$, she gets a detection. In this case, she recreates the reference-subcarrier pulse pairs scaling their powers up to make $P_\text{ref} = 3~\milli\watt$. When Bob also measures in the same basis as Alice-Eve and $\varphi_E = \varphi_B$ ($\varphi_E = \varphi_B \pm \pi$), this results in constructive (destructive) interference and will (will not) trigger a click in the blinded sideband detector. If Bob and Eve select different bases, Bob should not register any detection. For the slots when Eve gets no detection due to $\varphi_E \ne \varphi_A$ or low detection efficiency, she simply does nothing owing to the absence of reference pulse monitoring, and these events will appear as loss to Bob.

In order to successfully perform this attack in practice, the blinded detector should be characterized to know $P_\text{never}$ and $P_\text{always}$, which are the thresholds of the trigger pulse power making the detector never click and always click. The trigger pulse power $P_\text{tr}$ in the successful attack needs to satisfy the conditions
\begin{equation}
\begin{split}
P_\text{tr} \geq P_\text{always},\\
\frac{1}{2}P_\text{tr} \leq P_\text{never}.
\end{split}
	\end{equation} 
However, if the reference monitoring is implemented, both the reference and sideband monitoring detectors would be blinded and Eve will need to modify her strategy. When she has a conclusive outcome, she proceeds as before. However, when her measurement outcome is inconclusive, i.e.,\ $\varphi_E \ne \varphi_A$, Eve needs to tailor the power of the reference signal in such a way that it is enough to force a click on the blinded reference detector but not on the subcarrier detector.

Note that, sometimes Eve gets a detection when measuring in the opposite basis to that of Alice and has no way to know if her measurement result coincides with Alice's bit. However, these states are either not detected at Bob due to Eve-Bob basis mismatch or detected and then discarded during sifting due to Alice-Bob basis mismatch.

\section*{A more general attack than CBS attack mentioned in Ref.~\onlinecite{miroshnichenko2018}}
\label{sec:powerful-attack}

\begin{itemize}
	\item For each quantum signal going from Alice to Bob, Eve splits off a tiny fraction $x$ of each signal in the channel.
	\item Eve performs a quantum non-demolition measurement on the split signal~\cite{dusek2000}.
	\item If no photons are found, she splits off another fraction $x$. She does this until her induced loss equals the line loss.
	\item When photons are found, she keeps them in her quantum memory and sends the rest of the radiation to Bob via a lossless channel. The state of each photon in her possession is
	\begin{equation}
	\ket{\psi}_e   = (a^\dagger + m e^{i\varphi_A} b^\dagger) \ket{0}_A \ket{0}_B,
	\end{equation}
	where $a^\dagger$ and $b^\dagger$ are the creation operators on the carrier and subcarrier modes respectively, $m$ is the modulation index and $\varphi_A$ is Alice's phase encoding.
	\item For different values of $\varphi_A \in \{0, \pi/2, \pi, 3\pi/2\}$, Eve's states are not orthogonal. To make them orthogonal to each other, Eve needs to apply a filtering operation
	\begin{eqnarray}
	\begin{aligned}
	& A_\text{success} = m \ket{0}_B \bra{0}_B + \ket{1}_B \bra{1}_B,\\
	& A_\text{fail} = I - A_\text{success}.
	\end{aligned}
	\end{eqnarray}
	This turns $\ket{\psi}_e$ into
	\begin{equation}
	\ket{\psi'}_e = (a^\dagger + e^{i\varphi_A} b^\dagger) \ket{0}_A \ket{0}_B
	\end{equation}
	with a success probability
	\begin{equation}
	P_\text{success} = \frac{2m^2}{1 + m^2}.
	\end{equation}
	\item When the bases are revealed during sifting, Eve simply measures $\ket{\psi'}_e$ in the correct basis to extract $\varphi_A$.	
\end{itemize}

This attack is more powerful because, in Ref.~\onlinecite{miroshnichenko2018}, for a line loss $\eta$ Eve uses a $(1-\eta) : \eta$ beam splitter and the attack only succeeds when both Eve and Bob receive a photon. This becomes less likely as the line loss increases. However, in the present case Eve is not restricted to split in the $(1-\eta) : \eta$ ratio for the line loss $\eta$, which gives her more power. Thus, the security proof should be updated to include more (and ideally the most) general attacks than the collective beam splitting attacks.

\section*{Asymptotic security}
\label{sec:asymptotic-proof}

\noindent
We assume here, that the family of protocols considering in this paper belongs to the class of one-way QKD protocols with  independent and identically-distributed (i.i.d.)\ information carriers and direct reconciliation. It is commonly accepted that secure key generation rate $K$ for the protocols of this class in the presence of collective attacks in asymptotic regime is lower bounded according to \cite{scarani2009,pirandola2020} by the Devetak-Winter bound \cite{devetak2005}
\begin{equation}\label{K}
K=\nu_S P_B \left[1-\mathrm{leak}_\text{EC}(Q)-\max_E\chi(A:E)\right],
\end{equation}
where $\nu_S$ is the repetition rate; $P_B$ is the probability of successful decoding and accepting a bit in a single transmission window; $Q$ is the quantum bit error rate (QBER), the probability that a bit accepted by Bob is erroneous; $\mathrm{code}_\text{EC}(Q)$ is the amount of information revealed by Alice through the public channel for the sake of error correction, which depends on QBER and is limited by the Shannon bound: $\mathrm{code}_\text{EC}(Q)\ge h(Q)$ where $h(Q)=-Q\log_2Q-(1-Q)\log_2(1-Q)$ is the binary Shannon entropy. Quantity $\chi(A:E)$ in \cref{K}) is the Holevo capacity, giving an upper bound for amount of information accessible to eavesdropper Eve in a given collective attack (quantum channel). It is well-known that coherent attacks in i.i.d.\ case can be bounded with collective attacks. So one usually considers coherent attacks as general collective attacks \cite{christandl2009postselection} in terms of arbitrary unitary operations on purified states in enlarged Hilbert space (described in terms of isometry) provided by Eve.

In Reference~\onlinecite{kozubov2019}, the result of arbitrary isometry is considered in order to estimate Holevo capacity in complementary channel. Eve performs unitary operation (described by isometry) between states in the channel and Eve's ancillas to make them (in general case) entangled in some way \cite{ekert1994eavesdropping}. It has been shown that Holevo capacity of complementary channel is maximized when states become untangled (but interacted). Further considering the property of isometry, i.e.,\ preserving the overlap between the states, it has been shown that highest mutual information between Alice and Eve is bounded by the Holevo bound. This statement eliminates the necessity to consider particular kinds of isometries.

In case of subcarrier-wave quantum key distribution Holevo bound can be found considering reduced unconditioned channel density operator, i.e.,\ considering only two states since Eve can wait to measure her states after reconciliation. Therefore the obtained Holevo bound using binary Shannon entropy function $h(x)=-x\log_2 x-(1-x)\log_2(1-x)$ of the unconditioned channel density operator eigenvalues is as follows:
\begin{equation}\label{chi2}
\chi(\rho) = h\left(\frac12(1-\exp\left[-\mu_0 \left(1-d^S_{00}(2\beta)\right)\right]\right),
\end{equation}
where $\mu_0$ is the amplitude of the coherent state on the of carrier wave determined by the average number of photons in a transmission window provided with coherent monochromatic light beam with optical frequency $\omega$, $d^S_{00}(\beta)$ is the Wigner d-function from the quantum theory of angular momentum \cite{varshalovich1988quantum}, and $\beta$ is determined by the modulation index $m$ \cite{miroshnichenko2018}.

\section*{Advanced unambiguous\hyp{}state\hyp{}discrimination attack}
\label{sec:usd-proof}

\noindent
The collective attack that considers a mutual information between Alice and Eve might not be the most general attack. There might be attacks that decrease conditional mutual information $I(A;B|E)$ to zero. An example of such attack has been introduced in Ref.~\onlinecite{gaidash2018overcoming} where Eve performs an errorless USD measurement \cite{peres1998optimal,chefles1998unambiguous} then blocks inconclusive results and alters (amplifies and adds errors) the distinguished states. The latter is necessary to maintain both detection and error rates. In Reference~\onlinecite{gaidash2018overcoming} the condition of revealing Eve's actions \cite{tamaki2010} is generalised as
\begin{equation}
P_\text{det}>P_\text{USD},
\end{equation}
where $P_\text{det}$ is an expected detection probability and $P_\text{USD}$ is the probability of unambiguous state discrimination. Obviously there are two main strategies to increase the performance of the system. The first is to increase $P_\text{det}$ and the second is to decrease $P_\text{USD}$. We refer to Refs.~\onlinecite{gaidash2019methods,gaidash2018overcoming} for a further discussion of proposed approaches against the USD attack.

\section*{Finite-key security}
\label{sec:finite-key-proof}

\noindent
Since the resources such as time and memory are finite, it is not sufficient to consider asymptotic security. Therefore, in Ref.~\onlinecite{kozubov2019} a finite-key analysis has been performed. To estimate appropriate bound on secure key rate we consider the notation of Renyi entropies $H_{\alpha}(X)=\frac{1}{1-\alpha}\log\left(\sum_{i=1}^n p_i^\alpha\right)$, because they describe the worst case and not the average one. In the paper we consider that $\alpha\rightarrow\infty$ since we use \textit{min-entropy} $H_\infty(X)=\entmin =-\log \max_i p_i$. Thereby a quantum asymptotic equipartition property (QAEP) \cite{tomamichel2009fully} is considered in order to bound \emph{$\varepsilon$-smooth min-entropy} by von Neumann entropy. It means that for a large number of rounds, the operationally relevant total uncertainty can be well approximated by the sum over all i.i.d.\ rounds. In SCW QKD, conditional von Neumann entropy, or more precisely an entropy of Alice's bit conditioned on Eve's side-information in a single round, is bound as $H(\bitstring{A}|\mathbf{E})\geq 1-\chi(\rho)$.

To provide the key extraction one should carry out the following steps.

(i) \textit{Parameter estimation.} One should estimate the error rate (Bob publicly sends a random subset of $k$ bits to Alice, and she estimates the QBER $Q_\text{est}$ in that subset) and detection rate at Bob's side.

(ii) \textit{Error correction.} At this step both legitimate parties should check and correct the errors in their bit strings. It can be done using any error correction code. 

(iii) \textit{Privacy amplification.} In Reference~\onlinecite{kozubov2019}, the privacy amplification has been studied using the bound from Ref.~\onlinecite{tomamichel2011leftover}, which tells us that the trace distance $d$ between the protocol's output and an ideal output (where the key is uniform and independent from Eve, even after Eve knows the matrix used for the hashing) is bound above by  
\begin{equation}
\begin{split}
d &= \frac{1}{2}\Vert \rho_\text{KFE}-\omega_{K}\otimes\sigma_\text{FE}\Vert_{1} \\
&\leq \varepsilon_s + \frac{1}{2}\sqrt{2^{l - \entmin^{\varepsilon_s}(\bitstring{A}'|\mathbf{E})}} \\
&\leq \varepsilon_s + \frac{1}{2}\sqrt{2^{-\lossPA}} \label{eq_epssec} \\
&\leq \varepsilon_s + \eps{}{PA} =\varepsilon_\text{sec},
\end{split}
\end{equation}
where in the last step the quantity $\varepsilon_{sec}$ is introduced as an upper bound on $d$.

Reference~\onlinecite{kozubov2019} gives the final result that the protocol is $\varepsilon_\text{corr}$-correct with $\varepsilon_\text{corr} = \eps{}{EC}$ and $\varepsilon_\text{sec}$-secure with $\varepsilon_\text{sec} = \varepsilon_s + \eps{}{PA}$, hence $\varepsilon_\text{QKD}$-secure-and-correct, with $\varepsilon_\text{QKD} = \eps{}{EC} + \varepsilon_s + \eps{}{PA}$ providing secure bit string with length
\begin{equation}\label{lqkd}
\begin{split}
l = n(1-\chi(\rho))-4\sqrt{n}\log(2+\sqrt{2})\sqrt{\log\left( \frac{2}{\varepsilon_S^2} \right)}- \\
-k-\text{code}_\text{EC}(Q)-\log\frac{1}{\eps{}{EC}}-\log\frac{1}{\eps{}{PA}}+2.
\end{split}
\end{equation}



\section*{Acknowledgments}
\noindent We thank M.~Legr{\' e}, N.~L{\" u}tkenhaus, E.~Tan and  R.~Renner for discussions. This work was funded by NSERC of Canada (programs Discovery and CryptoWorks21), MRIS of Ontario, and the Ministry of Education and Science of Russia (programs 5-in-100 and NTI center for quantum communications). P.C.\ was supported by Thai DPST scholarship. A.H.\ was supported by China Scholarship Council, the National Natural Science Foundation of China (grant 61901483), and the National Key Research and Development Program of China (grant 2019QY0702). H.Q.\ is sponsored by Shanghai Pujiang Program. This work was funded by Government of Russian Federation (grant 08-08).

\section*{Author contributions}
S.S.,\ P.C.,\ A.H.,\ H.Q.,\ and V.M.\ performed the security analysis. V.E.,\ A.K.,\ A.Ga.,\ V.C.,\ A.V.,\ and A.Gl.\ developed the SCW QKD system, its security proof and countermeasures to the attacks listed in the security analysis. All authors wrote the Article.

\section*{Competing financial interests}
The authors declare no competing financial interests.

\end{document}